\documentclass[hyper]{JHEP} 

\usepackage{epsfig}

\newcommand\fverb{\setbox\pippobox=\hbox\bgroup\verb}
\newcommand\fverbdo{\egroup\medskip\noindent%
			\fbox{\unhbox\pippobox}\ }
\newcommand\fverbit{\egroup\item[\fbox{\unhbox\pippobox}]}
\newbox\pippobox
\title{Note on  D-Brane Effective Action
in the Linear Dilaton Background}
\author{by J. Kluso\v{n}\\
	 Department of Theoretical Physics and Astrophysics\\
                   Faculty of Science, Masaryk University\\
Kotl\'{a}\v{r}sk\'{a} 2, 611 37, Brno\\
Czech Republic\\
	E-mail: \email{klu@physics.muni.cz}}

\preprint{\hepth{0310066}}

\abstract{In this short note we will study 
effective action for unstable D-brane in
the linear dilaton background. We will solve
the equation of motion for large $T$ and we will
calculate the stress energy tensor. Then we compare
our results with the calculations performed using
exact conformal field theory description of the open
string worldsheet theory.}

\def\bb{\mathbf{B}}

\def\ss{\sin \frac{\tau}{\sqrt{2}}}
\def\st{\sinh \frac{\tau}{\sqrt{2}}}
\def\st2{\sinh^2 \frac{\tau}{\sqrt{2}}}
\def\ss2{\sin^2 \frac{\tau}{\sqrt{2}}}

\begin{document}
\section{Introduction}\label{first}
The study of time-dependent open string
tachyons \cite{Sen:2002nu,Sen:2002in,Sen:2002vv,Sen:2002qa,
Strominger:2002pc,Gutperle:2002ai,Gutperle:2003xf}
was very intensive in recent two
years and also leads to the discovery 
of new dualities and reinterpretation some
old ones, especially relation between
two dimensional string theories and
matrix models  \cite{Larsen:2002wc,Maloney:2003ck,Okuda:2002yd,
Sugimoto:2002fp,McGreevy:2003kb,Klebanov:2003km,
Lambert:2003zr,Gaiotto:2003rm,McGreevy:2003ep,
Karczmarek:2003xm,Sen:2003iv,Gutperle:2003ij,
Kluson:2003sh,Sugawara:2003xt,Sen:2003xs,
Kluson:2003rd,Constable:2003rc,Douglas:2003up,
Teschner:2003qk,Takayanagi:2003sm,
Nagami:2003yz,Fredenhagen:2003ut,Okuyama:2003jk,
Klebanov:2003wg,DeWolfe:2003qf,Karczmarek:2003pv,Giveon:2003wn,
Teschner:2003qk,Kapustin:2003hi}.

In the same way  recent works on dynamics
of unstable D-branes in string theory has
led to an effective action for the open string
tachyon $T$ and massless open string modes 
$A_{\mu}$ (the gauge field on the D-brane) and
$Y^I$ (the scalar field parametriziing the location
of the D-brane in the transverse directions) 
\cite{Sen:1999md,Garousi:2000tr,
Bergshoeff:2000dq,Kluson:2000iy,
Lambert:2001fa,Lambert:2002hk,
Garousi:2003pv,Kutasov:2003er,
Okuyama:2003wm}. This action has the
form 
\footnote{
We use the conventions $\alpha'=1 \ ,
\eta_{\mu\nu}=(-1,+1,\dots,+1).$}
\begin{eqnarray}\label{act}
S=\int d^{p+1}x \mathcal{L} 
\nonumber \\
\mathcal{L}=-\tau_pV(T)\sqrt{-\det \mathbf{A}} \ ,
\nonumber \\
\mathbf{A}_{\mu\nu}=
\eta_{\mu\nu}+
\partial_{\mu}T\partial_{\nu}T+
\delta_{IJ}\partial_{\mu}Y^I\partial_{\nu}
Y^J+F_{\mu\nu} \ , \nonumber \\ 
\end{eqnarray}
where $\tau_p$ is Dp-brane tension
and  $V(T)$ is tachyon potential that for
large $T$ is expected to behave as
\begin{equation}
V(T)\sim e^{-\alpha T/2} \ ,
\end{equation}
with $\alpha=1$ for a bosonic string and $\alpha=
\sqrt{2}$ for the non-BPS D-brane in
superstring. The action (\ref{act}) is known
to reproduce some  aspects of open string
dynamics. For example, if we choose 
 the tachyon potential as \cite{Lambert:2003zr}
\begin{equation}
V(T)=\frac{1}{\cosh\frac{\alpha T}{2}} \ ,
\end{equation}
one finds from  (\ref{act}) the correct stress
energy tensor $T_{\mu\nu}$ in homogeneous tachyon
condensation. For the case of an unstable
D-brane in Type II string theory we can also construct
a codimension one BPS D-brane as a solitonic
solution of (\ref{act}). 

Thanks to these results one can believe that 
the action (\ref{act}) captures some class
of phenomena of classical open string theory. 
Following \cite{Kutasov:2003er}
we can consider the action (\ref{act}) as generalisation
of the DBI action describing the gauge field
$A_{\mu}$ and scalars $Y^I$ on 
D-brane. The DBI action is valid in the full  open
string theory in situations when $F_{\mu\nu}$ 
and $\partial_{\mu}Y^I$ are arbitrary but
slowly varying.

Since the success of the action (\ref{act}) 
in the description
of the tachyon  condensation in the flat spacetime
is very intriguing it
seems to be natural to consider 
 this action in a more
general  closed string  background. Recently the
exact conformal field theory (CFT) analysis
of the
time-dependent tachyon condensation in
the linear dilaton background 
was performed in \cite{Karczmarek:2003xm}.
We mean that would be interesting to look at
this problem also from
the  effective action point of view. 
 The starting point of our calculation is
the presumption that the action (\ref{act})
gives good description of the effective field
theory dynamics of unstable D-brane 
in the  linear  
dilaton background $\Phi=V_{\mu}x^{\mu}$ in the situation when
the tachyon field $T$ is large.
Then  the equation of motion implies an asymptotic
behaviour of the tachyon as
  $T\sim \beta t$ where $\beta $ is 
constant which depends on  time component $V_0$
of the spacelike vector $V_{\mu}$. It turns out however
that the dependence of $\beta$ on 
$V_{\mu}$ is not as the same as the exact 
CFT relation 
$\beta(\beta-V_0)=1$
\cite{Karczmarek:2003xm}. We also find that the
asymptotic behaviour of the stress energy tensor derived
from (\ref{act})  is completely 
different from the exact calculation  given  in
\cite{Karczmarek:2003xm}. More precisely, we
will  find that the
stress energy tensor diverges at far future.  
These results imply that the effective action 
description of the tachyon condensation in the linear
background is much more complicated than in
case of the constant string coupling.

The organisation of this paper is as follows. In the
next section (\ref{second}) we review how to get
 the solution
of  the tachyon equation of motion on unstable
D-brane when $T$ is large and when
 the    dilaton field is constant. 
 Then we 
extend this analysis to 
the case of the
linear dilaton background. In order to understand
better of the problem of the tachyon condensation
in the linear dilaton background we will 
  study in section (\ref{third}) D-brane effective action
proposed in  \cite{Kutasov:2003er}. 
We will show that even if this action
correctly describes the mass of the tachyon fluctuation
around the unstable vacuum $T=0$ in case of
the constant dilation its description of the
tachyon dynamics in the linear dilaton background 
is more involved. In particular, we will show that
 on-shell condition for tachyon fluctuations around
unstable vacuum $T=0$ is different from the exact one.  
In conclusion (\ref{fourth}) we
outline our results and also suggest other problems
that deserve further study.

\section{D-brane effective action}\label{second}
In this section we will study the process of
the tachyon condensation on unstable D-brane
in the bosonic theory  when the closed string
background is nontrivial.
 As we claimed  in the introduction
the main point of our interest is the Dp-brane
effective action
\begin{eqnarray}\label{acte}
S=-\tau_p\int d^{p+1}xe^{-\Phi}
V(T)\sqrt{-\det \mathbf{A}}=
-\tau_p\int d^{p+1}x
e^{-\Phi}\sqrt{-g}V(T)\sqrt{\bb} 
 \ ,
\nonumber \\
 \mathbf{A}\equiv \eta_{\mu\nu}+
\partial_{\mu}T\partial_{\nu}T \ ,
g=\det \eta=-1 \ , 
\nonumber \\
-\det\mathbf{A}\equiv
-g\bb=\left(1+\eta^{\mu\nu}\partial_{\mu}T
\partial_{\nu}T\right) \ , \nonumber \\
\end{eqnarray}
where we restrict ourselves to the dynamics of
the  tachyon $T$ only. 
In this place it is important to stress that 
 the action (\ref{acte}) was not obtained from the first
principles, it is only consistent with the time
evolution of the stress energy tensor of various
classical solutions in string theory in late times
\cite{Sen:2002qa}. In the same way
 the form of the potential was derived
by requiring that during the rolling tachyon process
the pressure associated with this configuration
falls off as $e^{-t}$. 

Before we proceed to the study of 
 the tachyon dynamics
in the linear dilaton background we firstly review
the rolling tachyon solution 
 in case of the constant $\Phi$. 
For the  time-dependent tachyon $T(t)$  the equation 
of motion that arises from
(\ref{acte})  has the form
\begin{equation}\label{eqe}
e^{-\Phi}V'(T)\sqrt{\bb}+\frac{d}{dt}
\left(\frac{e^{-\Phi}\dot{T}}{\sqrt{\bb}}\right) =0 \ .
\end{equation}
Since for large $T$ the expected behaviour of $V$ is $V=e^{-\alpha
T/2}$ it is natural to consider following asymptotic form of $T$
\begin{equation}\label{ans}
\dot{T}=1-K^2e^{-\alpha t}  \ ,
\dot{T}^2=1-2K^2e^{-\alpha t}
\Rightarrow
\sqrt{\bb}=\sqrt{2}Ke^{-\alpha t/2} \ .
\end{equation}
It is easy to see that the ansatz (\ref{ans})
solves the equation of motion
(\ref{eqe})
\begin{eqnarray}
-\frac{\alpha }{2}e^{-\alpha T/2}
\sqrt{2}K e^{-\alpha t/2}
+\frac{d}{dt}
\left(\frac{e^{-\alpha t/2}\dot{T}}
{\sqrt{2} K e^{-\alpha t/2}}
\right)
=0 \Rightarrow \nonumber \\
-\frac{\alpha K}{\sqrt{2}}e^{-\alpha t}+
\frac{d}{dt}\frac{(1-K^2e^{-\alpha t})}{\sqrt{2}K}
=-\frac{\alpha K}{\sqrt{2}}e^{-\alpha t}+
\frac{\alpha K}{\sqrt{2}}e^{-\alpha t}=0 \ .
\nonumber \\
\end{eqnarray}
We must stress that
the tachyon  grows at far future as 
$T\sim \beta t  \ , \beta=1$. On the other hand in
the CFT description of the rolling tachyon the parameter
$\beta$ is presented in the exponential of the marginal
operator $T\sim e^{\beta X^0} \ ,  \beta=1$ which
is  inserted
on the boundary of the worldsheet.
 We then mean that it is reasonable to
ask the question whether there is similar  relation between
the asymptotic form  of tachyon field in the effective
theory description $T\sim \beta t$ and
the parameter $\beta'$ in the worldsheet boundary
interaction term $T\sim e^{\beta't}$ when the
string is embedded  in
the linear dilaton  
background 
\begin{equation}
\Phi=V_{\mu}x^{\mu} \ ,
V_{\mu}V^{\mu}=(26-D)/6>0 \ .
\end{equation}
More precisely, the
 CFT description of the tachyon condensation
on unstable D-brane in the linear 
dilaton background is as
follows \cite{Karczmarek:2003xm}.  
Tachyon condensates
can be in the form $\exp \left(\beta' X^0\right)$
 with 
$\beta' >0$. The conformal dimension 
of worldsheet 
boundary interaction operator is $\triangle=\beta(\beta'-V_0)$.
Requiring that the boundary interaction is marginal implies
that $\triangle=1$ and hence we obtain
\begin{equation}\label{exact}
V_0=\beta'-\frac{1}{\beta'} \ .
\end{equation}
The question is whether 
we can find 
 similar relation between 
$V_0$ and  
the parameter $\beta$ in the effective theory description
if  we propose  following
form of $T$ at far future
\begin{equation}\label{Tas}
\dot{T}=\beta-Ke^{-\gamma t} \ ,
\dot{T}^2=\beta^2-2K\beta e^{-\gamma t} \ ,
\sqrt{\bb}=\sqrt{1-\beta^2}+
\frac{K^2\beta}{\sqrt{1-\beta^2}}
e^{-\gamma t} \ .
\end{equation}
 We  presume 
 that $\gamma >0$ so that the exponential
term vanishes for large $t$. We also demand that
$\beta \neq 1$  in order 
$\sqrt{1-\beta^2}$ to be   finite. 
 Then the equation 
of motion is
\begin{eqnarray}\label{eq1}
-\frac{\alpha}{2}e^{-\Phi-\alpha \beta t/2}
\sqrt{1-\beta^2}+
\frac{d}{dt}\left(\frac{e^{-\Phi-\alpha \beta t/2 }
\beta}{\sqrt{1-\beta^2}}\right)=0 
\Rightarrow \nonumber \\
\Rightarrow -\frac{\alpha}{2}(1-\beta^2)
-V_0\beta -\frac{\alpha \beta^2}{2}=0
\Rightarrow \alpha=-2V_0\beta  \ . 
\nonumber \\
\end{eqnarray}
Since  the system rolls to
the stable vacuum at $T=\infty$ we  must have   $\beta >0$
and hence  
 (\ref{eq1}) implies   $V_0<0$. 
In other words the tachyon condensation when
the tachyon reaches its stable minimum at 
$T=\infty$ is only possible when the string
coupling constant vanishes at far future.
Since we are interested in the dynamics
of unstable D-brane in bosonic theory we have
$\alpha=1$ and  then from 
 (\ref{eq1}) we get
\begin{equation}\label{Vbeta}
V_0=-\frac{1}{2\beta}   \ .
\end{equation}
This result is clearly different from
the exact CFT relation (\ref{exact}).
We  mean that this is the first indication 
 that the effective field theory 
description of the tachyon condensation
in the linear dilaton background does not
completely reproduce results obtained
through 
CFT analysis. Further support for
this claim follows from the
 study of the asymptotic behaviour of
the stress energy
tensor $T_{\mu\nu}$ defined as
\begin{equation}
T_{\mu\nu}=-\frac{2}{\sqrt{-g}}
\frac{\delta S}{\delta g^{\mu\nu}} \ .
\end{equation}
From (\ref{acte}) and
using $\frac{\delta \sqrt{-q}}{\delta
g^{\mu\nu}}=-\frac{1}{2}\sqrt{-g}
g_{\mu\nu}$ 
we obtain 
\begin{eqnarray}\label{stressrol}
T_{\mu\nu}=-\eta_{\mu\nu}e^{-\Phi}V(T)
\sqrt{\bb}+\frac{e^{-\Phi}V(T)\partial_{\mu}T\partial_{\nu}T}
{\sqrt{\bb}} \ , \nonumber \\
T_{00}=\frac{e^{-\Phi}V(T)(1+\eta^{ij}
\partial_i T\partial_j T)}{\sqrt{\bb}} \ , \nonumber \\
T_{ij}=-\delta_{ij}\frac{e^{-\Phi}V(T)}{\sqrt{\bb}}
(1-\partial_0 T\partial_0 T) \ . \nonumber \\
\end{eqnarray}
Hence for solution  (\ref{Tas}) components of
the stress energy tensor at far future are equal to
\begin{eqnarray}\label{stressrola}
T_{00}=\frac{e^{\left(\frac{1-4V_0^2}{4V_0}\right)t}}
{\sqrt{1-\beta^2}}
 \ ,\nonumber \\
T_{ij}=-\delta_{ij}
e^{\left(\frac{1-4V_0^2}{V_0}\right)t}
\sqrt{1-\beta^2}
 \  .
\nonumber \\
\end{eqnarray}
From the requirement  that the stress
energy tensor should be real and finite 
we get  the condition
$\beta^2<1$ which  using
(\ref{Vbeta}) also implies 
\begin{equation}
V_0>-1/2 \ .
\end{equation}
From (\ref{stressrola}) we also see 
that for $-1/2<V_0<0$ all components
of the stress energy tensor
diverge at far future. This behaviour
is completely different  from 
 the exact CFT description given in 
\cite{Karczmarek:2003xm} where it was shown
that zero component of the stress energy tensor 
$T_{00}$ does not depend on time at far future while
the spatial components $T_{ij}$ vanish.
We mean that these results show that
the   effective
action (\ref{act})  description of
the tachyon condensation in the linear
dilaton background is slightly problematic. 
On the other hand it is possible that our proposed ansatz
(\ref{Tas}) is too naive and more complicated
form of the tachyon at far future could give better
results. In any case we see that the effective
action description of the tachyon condensation in
the linear dilaton background is much more complicated
than in case of constant $\Phi$.
More comments about this result will be given
in conclusion. 

In the next section 
we will study the tachyon condensation
in the linear dilaton background using
an   effective action that was proposed
\cite{Kutasov:2003er}.
\section{Other form of the D-brane effective
action }\label{third}
We begin this section with the
brief  review how the D-brane
effective action in 
bosonic string theory was derived
in \cite{Kutasov:2003er}. 
We start with the D-brane
effective action proposed 
in  \cite{Lambert:2003zr}
\begin{equation}\label{org}
S=-\tau_p\int e^{-\Phi}\frac{1}{
\cosh \frac{\tilde{T}}{2}}\sqrt{1+\eta^{\mu\nu}\partial_{\mu}
\tilde{T}\partial_{\nu}\tilde{T}} \ .
\end{equation}
Using the following  field redefinition  
\cite{Kutasov:2003er} 
\begin{equation}\label{Kutfield}
T=\sinh^2\frac{\tilde{T}}{2} \ , \
\dot{T}=\dot{\tilde{T}}\cosh \frac{\tilde{T}}{2}
\sinh \frac{\tilde{T}}{2}
\end{equation}
we obtain an action in the form
\begin{equation}\label{Kutbos}
S=-\tau_p\int e^{-\Phi}
\frac{1}{1+T}\sqrt{1+T+\frac{\eta^{\mu\nu}
\partial_{\mu}T\partial_{\nu}T}{T}} \ .
\end{equation}
We must say few words  about the action
(\ref{Kutbos}). The main point is that the Lagrangian
in (\ref{org}) is analytic in $\tilde{T}$ while the
Lagrangian in (\ref{Kutbos}) is non-analytic in
terms of $T$. This result has important consequence
and we mean that this is the reason why the
effective action (\ref{Kutbos}) deserves to be studied
 separately.
In particular, it was shown that the
 action  (\ref{org}) does not give the correct
mass of the tachyon fluctuations around
the unstable minimum $\hat{T}=0$. 
As was explained in \cite{Kutasov:2003er} this
disagreement can be understood to be due to
the non-analytic relation (\ref{Kutfield}) between
the open string tachyon $T$ and the field $\tilde{T}$
that appears in (\ref{org}). Taking the map into
account we can show that (\ref{Kutbos}) correctly
describes on-shell physics of the tachyon
fluctuations around the unstable minimum $T=0$.
 More precisely, let us expand 
(\ref{Kutbos}) around unstable minimum of
the potential  $T=0$ when we presume that
$T, \frac{\eta^{\mu\nu}\partial_{\mu}T
\partial_{\nu}T}{T}$ are small
\begin{equation}\label{actlin}
S=-\tau_p\int 
e^{-\Phi}\left(
\frac{\eta^{\mu\nu}\partial_{\mu}T
\partial_{\nu}T}{2T}-\frac{T}{2}\right) \ .
\end{equation}
Then the equation of motion that
arises from (\ref{actlin}) is 
\begin{equation}\label{kuteq}
-\frac{e^{-\Phi}}{2}
-\frac{e^{-\Phi}\eta^{\mu\nu}
\partial_{\mu}T\partial_{\nu}T}{2T^2}
-\partial_{\mu}\left(
\frac{e^{-\Phi}\eta^{\mu\nu}
\partial_{\nu}T}{T}\right)=0 \ . 
\end{equation}
For constant $\Phi$ it is easy to see that the plane wave ansatz
\begin{equation}\label{plane}
T=e^{ik_{\mu}x^{\mu}}
\end{equation}
 is solution of the equation of motion
on the condition
\begin{equation}
-1+k_{\mu}\eta^{\mu\nu}k_{\nu}=0  \ 
\end{equation}
which is correct on shell condition for the tachyonic mode
in bosonic theory. However for the linear dilaton background
we obtain disagreement with the exact CFT condition. 
The origin of this result is the fact that for the
plane wave ansatz (\ref{plane}) 
the third term in (\ref{kuteq}) is nonzero for linear
dilaton background and hence the mass shell condition is
 \begin{equation}
-1+k_{\mu}\eta^{\mu\nu}k_{\nu}
+2iV_{\mu}\eta^{\mu\nu}k_{\nu}=0 \ .
\end{equation}
On the other hand the exact CFT result gives 
 \begin{equation}
-1+k_{\mu}\eta^{\mu\nu}k_{\nu}
+iV_{\mu}\eta^{\mu\nu}k_{\nu}=0 \ .
\end{equation}
We suspect that some corrections to the effective
actions that are proportional to the derivatives
of $\Phi$ become important for description
of the effective action in the linear dilaton background,
since for constant $\Phi$ the effective
action (\ref{Kutbos}) is in very good agreement
with the exact CFT description.
However to find these additional terms in the effective
action is very difficult task and it is behind
the scope of this paper. 

Next we would like to solve the equation of motion for
the time-dependent tachyon solution in case of large $T$.
If we denote $\bb=1+T+\frac{\eta^{\mu\nu}
\partial_{\mu}T\partial_{\nu}T}{T}$ then 
the equation of motion that arises from
(\ref{Kutbos})  has
the form
\begin{equation}
-\frac{e^{-\Phi}}
{(1+T)^2}\sqrt{\bb}+\frac{e^{-\Phi}}{2(1+T)\sqrt{\bb}}
-\frac{e^{-\Phi}\eta^{\mu\nu}\partial_{\mu}T
\partial_{\nu}T}{2(1+T)T^2\sqrt{\bb}}-
\partial_{\mu}\left(\frac{e^{-\Phi}\eta^{\mu\nu}
\partial_{\nu}T}{(1+T)T\sqrt{\bb}}\right)=0 \
\end{equation}
that for large $T$ reduces to
\begin{equation}\label{eqlarg}
-\frac{e^{-\Phi}}{T^2}\sqrt{\bb}+\frac{e^{-\Phi}}{2T\sqrt{\bb}}
-\frac{e^{-\Phi}\eta^{\mu\nu}\partial_{\mu}T
\partial_{\nu}T}{2T^3\sqrt{\bb}}-
\partial_{\mu}\left(\frac{e^{-\Phi}\eta^{\mu\nu}
\partial_{\nu}T}{T^2\sqrt{\bb}}\right)=0 \ ,
\end{equation}
where now $\bb=\frac{T^2-\dot{T}^2}{T}$. For the ansatz
\begin{equation}\label{kutans}
T=ke^{\beta t}  \ 
\end{equation}
we get
 \begin{equation}
\dot{T}=\beta T \  ,
\bb=T(1-\beta^2) \ .
\end{equation}
Inserting these expressions into 
(\ref{eqlarg}) we finally obtain
\begin{eqnarray}
-\frac{e^{-\Phi}\sqrt{1-\beta^2}
}{T^{3/2}}+\frac{e^{-\Phi}}{2T^{3/2}\sqrt{1-\beta^2}}
+\nonumber \\
+\frac{e^{-\Phi}\beta^2}
{2T^{3/2}\sqrt{1-\beta^2}}+
\frac{d}{dt}\left(\frac{e^{-\Phi}\beta }{
T^{3/2}\sqrt{1-\beta^2}}\right)=0
\Rightarrow \nonumber \\
\Rightarrow
-1+3\beta^2-2\beta V_0
-3\beta^2=0 
\Rightarrow \beta=-\frac{1}{2V_0}  \ . 
\nonumber \\
\end{eqnarray}
We see that we have got the same relation
between   
$\beta$ and $V_0$ as in
(\ref{Vbeta}).
 In the same
way as in the previous section
we can calculate from (\ref{Kutbos}) the stress
energy tensor that now is equal to
\begin{eqnarray}
T_{\mu\nu}=
-\frac{
e^{-\Phi}\eta_{\mu\nu}\sqrt{\bb}}{1+T}
+\frac{e^{-\Phi}\partial_{\mu}T
\partial_{\nu}T}{(1+T)T\sqrt{\bb}} \ , \nonumber \\
T_{00}=\frac{e^{-\Phi}}{\sqrt{1+T-\frac{\dot{T}^2}{T}}} \ ,
\nonumber \\
T_{ij}=-\frac{\delta_{ij}e^{-\Phi}\sqrt{1+T-\frac{\dot{T}^2}{T}}}
{1+T} \ . \nonumber \\
\end{eqnarray}
Then for the ansatz  (\ref{kutans}) we
get 
\begin{eqnarray}
T_{00}=\frac{e^{\frac{1-4V_0^2}{4V_0}t}}{
\sqrt{1-\beta^2}} \ , \nonumber \\
T_{ij}=-\delta_{ij}e^{\frac{1-4V_0^2}{4V_0}t}
\sqrt{1-\beta^2} \ . \nonumber \\
\end{eqnarray}
As we could expect the asymptotic form of the stress
energy tensor is the same as in the previous section
since non-analycity of the field redefinition (\ref{Kutfield})
does not show  for large $T$.
In spite of this fact 
we  mean that it was useful to perform the analysis of
the effective D-brane action
(\ref{Kutbos}) in the linear dilaton background too. 
In particular, using
the action (\ref{Kutbos}) we were able
to study the behaviour of the tachyon fluctuations
around the unstable minimum $T=0$. We have shown
that even 
in this region  the
effective action (\ref{Kutbos}) 
 does not correctly reproduces exact CFT on shell
condition for 
 the tachyonic fluctuations  
in the linear dilaton background. On the contrary
 the action
(\ref{Kutbos}) is very successful in the description
of the tachyon dynamics on the world volume of
unstable D-brane when the dilaton field is
constant. 
We mean that this result again suggests that the 
effective field theory description of the unstable D-brane
in the linear dilaton background is rather subtle
and much more complicated than in case of constant
dilaton.
\section{Conclusion}\label{fourth}
In this paper we have considered the effective
field theory description of an unstable D-brane
in the linear dilaton background. We have asked
the question whether the effective
field theory description of
the time-dependent 
process of the tachyon condensation
in this background can reproduce exact
CFT calculation performed in 
\cite{Karczmarek:2003xm}.  
 Since it is generally
believed that the effective action
(\ref{act}) is valid for large tachyon $T$  
\cite{Sen:2002qa} we restrict ourselves to 
the study of the asymptotic behaviour of $T$ at
far future where the tachyon field approaches 
its stable minimum at $T=\infty$.
 We have begun with the action (\ref{act}) 
which reproduces correctly the stress energy tensor
for rolling tachyon solution
\cite{Sen:2002nu,Sen:2002qa,
Lambert:2003zr,Lambert:2002hk,Kutasov:2003er}
in case of the constant dilaton field. 
 We have found solution
of equation of motion for large $T$ and we have calculated
the stress energy tensor. According to our results the
tachyon condensation on unstable D-brane in
the  bosonic theory
 can occurs in case when the string
coupling constant vanishes at far future. 
This is different from the exact CFT description where
the tachyon condensation occurs for any 
spacelike dilaton vector. 
Then we have calculated the stress energy tensor
for the rolling tachyon solution and
we have found that all  its components 
 diverge. On the other hand CFT analysis
\cite{Karczmarek:2003xm} 
showed that for $t\rightarrow \infty$ $, T_{00}$ 
is independent on time while $T_{ij}$ decays
to zero which was interpreted as an emergence of
the tachyon dust at far future. As was argued 
there the fact, that despite the time dependence
of the string coupling, the energy density 
becomes constant and the pressure goes to 
zero suggests that the tachyon dust is a gas
of massive closed strings whose energy is not
affected by a change in the dilaton the
way the energy of a D-brane is. On the other
hand in the effective field theory description
all components of the stress energy tensor go to
infinity  
and there is no interpretation of the remnant of
the tachyon condensation as pressureless dust. 
We can  explain the divergence of the stress 
energy tensor as a consequence of the fact that
the stress energy tensor is proportional to $\sim g_s^{-1}=
e^{-\Phi}$ and according to the effective field theory 
analysis the tachyon condensation can occur only
in case when the string coupling constant vanishes at
far future. 
We mean that in order to describe the tachyon
condensation in the linear background using
the D-brane effective action  we should include
terms that are proportional to the derivation
of the dilaton and hence vanish when it is constant.
There is also possibility that in order to 
correctly describe D-brane in the linear dilaton
background   we should
include  the coupling to 
the closed string tachyon 
  into the D-brane effective action in order to
restore well known ``Liouville wall''. In any case,
we mean that the effective action description
of the tachyon condensation is very interesting
and reach subject in the string theory and 
there are many open problems that deserve further
study. 
\\
\\
{\bf Acknowledgement}

This work was supported by the
Czech Ministry of Education under Contract No.
14310006.
\\
\\

\end{document}